\documentclass[superscriptaddress,a4paper,tightenlines,nofootinbib,floatfix,11pt]{article}
\usepackage{jheppub}

\usepackage[T1]{fontenc}

\usepackage{setspace}
\usepackage{multirow,array}
\usepackage{amsmath,amssymb}
\usepackage{graphicx}
\usepackage{color}
\usepackage{url}
\usepackage{feynmp}
\usepackage{slashed}
\usepackage{multirow}
\usepackage{footmisc}
\usepackage{hyperref}
\usepackage[scientific-notation=true]{siunitx}
\usepackage{dsfont}
\usepackage{ulem}
\usepackage{cancel}
\usepackage{tikz}
\usepackage{soul}
\usepackage[compat=1.0.0]{tikz-feynman} 
\usepackage{arydshln}

\tikzfeynmanset{
  every dot={/tikz/fill=red!30,/tikz/inner sep=1.5pt},
}

\newcommand{\om}{\omega^{2/3}}
\newcommand{\omstar}{\omega^{-2/3}}
\newcommand{\xone}{X_{1}}
\newcommand{\xonestar}{X_{1}^*}
\newcommand{\xtwo}{X_{2}}
\newcommand{\xtwostar}{X_{2}^*}

\hypersetup{
   colorlinks=true,       
   linkcolor=blue,        
   citecolor=red,         
   filecolor=magenta,      
   }
   
\title{Lepton Number Violation at the LHC in Radiative Neutrino Mass Models with Leptoquarks}

\author[1]{K.S. Babu,}
\author[2]{Rahool Kumar Barman,}
\author[1]{and Dorival Gon\c{c}alves}
\affiliation[1]{Department of Physics, Oklahoma State University, Stillwater, OK, 74078, USA}
\affiliation[2]{Kavli Institute for the Physics and Mathematics of the Universe (WPI), The University of Tokyo Institutes for Advanced Study, The University of Tokyo, Kashiwa, Chiba 277-8583, Japan}

\emailAdd{babu@okstate.edu}
\emailAdd{rahoolbarman@physics.iitm.ac.in}
\emailAdd{dorival@okstate.edu}

\abstract{We investigate the prospects for observing lepton number violation (LNV) by two units, $|\Delta L| = 2$, at the LHC within the leptoquark variant of the Zee Model, where Majorana neutrino masses arise radiatively at one-loop. The model features an $SU(2)_L$ doublet and singlet leptoquarks, whose interactions produce a distinctive same-sign dilepton plus jets signature, $pp \to \ell^{\pm}\ell'^{\pm} + \text{jets}$. Taking into account current experimental constraints, we identify the dominant production channels for this LNV signal and perform a detailed collider analysis. We find that the HL-LHC can probe leptoquark masses up to $m_{\rm LQ} \sim 1.5~\mathrm{TeV}$ with this process. Observation of this signal would provide a direct test of LNV and would unambiguously establish the Majorana nature of neutrinos.}
\begin{document}

\maketitle

\section{Introduction}
\label{sec:intro}

Majorana neutrino masses can naturally arise from loop-induced processes, and a wide range of radiative mechanisms has been proposed in the literature at one-loop~\cite{Zee:1980ai,Hall:1983id}, two-loop~\cite{Zee:1985id,Babu:1988ki,Carquin:2019xiz}, and three-loop levels~\cite{Krauss:2002px}. For reviews, see Refs.~\cite{Cai:2017mow,Babu:2019mfe}. Among the minimal one-loop realizations is the Zee Model~\cite{Zee:1980ai}, which, due to the electroweak $s$-channel production of its charged scalars, offers promising sensitivity to lepton number violating (LNV) signature at the high-luminosity LHC (HL-LHC)~\cite{Babu:2022ycv}.

A central theoretical implication of observing LNV at colliders is rooted in the so-called {\it black-box theorem} for neutrino mass~\cite{Schechter:1981bd}. Originally formulated in the context of neutrinoless double beta decay ($0\nu\beta\beta$)~\cite{PhysRev.56.1184}, the theorem states that any process that violates lepton number by two units, $|\Delta L| = 2$, necessarily induces a nonzero Majorana mass for neutrinos, irrespective of the underlying mechanism. For a recent evaluation of the quantitative impact of the theorem, see Ref.~\cite{Duerr:2011zd}.
The theorem can be generalized to high-energy collider processes~\cite{Babu:2022ycv}, establishing that the observation of a $|\Delta L| = 2$ final state at the LHC would likewise constitute direct evidence for the Majorana nature of neutrinos. Unlike $0\nu\beta\beta$ searches, which only probe effective LNV interactions with two electrons $e^-e^-$, the LHC can  access effective interactions with electrons and muons in multiple combinations: $e^\pm e^\pm$, $\mu^\pm\mu^\pm$, and $e^\pm\mu^\pm$. This greatly broadens the scope of experimental tests of LNV, providing complementary and, in some cases, unique avenues to probe the Majorana nature of neutrinos.

The LNV signal $pp \rightarrow \ell^\pm \ell'^\pm + \text{jets}$ has been extensively studied in several theoretical frameworks. This includes the Left-Right Symmetric Model (LRSM)~\cite{Keung:1983uu,Gninenko:2006br,Maiezza:2010ic,Nemevsek:2011hz,Chen:2011hc,Chakrabortty:2012pp,Aguilar-Saavedra:2012grq,Han:2012vk,Chen:2013foz,Dev:2013wba,Dutta:2014dba,Gluza:2015goa,Ng:2015hba,Maiezza:2015lza,Deppisch:2015qwa,Degrande:2016aje,Dev:2016dja,Roitgrund:2017byx,Nemevsek:2018bbt}, type-I seesaw framework~\cite{Dicus:1991fk,Datta:1993nm,Ali:2001gsa,Han:2006ip,Kersten:2007vk,delAguila:2007qnc,Atre:2009rg,Alva:2014gxa,Das:2015toa,Das:2016hof,Das:2017gke,Drewes:2019byd,Fuks:2020att}, and more recently type-II seesaw and Zee Models~\cite{Babu:2022ycv,Bolton:2024thn}. 
The complementarity between neutrinoless double beta decay and LNV searches at the LHC with $pp \rightarrow \ell^\pm \ell'^\pm + \text{jets}$ has also been investigated in simplified model frameworks~\cite{Helo:2013dla,Helo:2013ika,delAguila:2013yaa,delAguila:2013mia,Peng:2015haa,Harz:2021psp,Graesser:2022nkv}. Experimental searches for the same-sign dilepton plus jets signature have been performed by ATLAS~\cite{ATLAS:2018dcj} and CMS~\cite{CMS:2018jxx} in the context of the LRSM and the Type-I seesaw, respectively. Related new physics signatures have also been studied at future lepton colliders~\cite{Rizzo:1981dla,Crivellin:2018ahj,Li:2023lkl,Mikulenko:2023ezx,Maharathy:2023dtp,deLima:2024ohf,Cao:2024rzb,Yang:2025jxc}, offering a complementary perspective to the same-sign dilepton plus jets channel searches at the LHC. Complementary studies have also been carried out at LHCb~\cite{LHCb:2014osd}, focusing on LNV signal from $B$-meson decays.

Motivated by the interplay between neutrino mass generation and collider observables, we investigate the leptoquark analog of the Zee Model in which the original scalar fields are replaced by leptoquarks. This variant preserves the radiative origin of Majorana neutrino masses while benefiting from the sizable QCD production of leptoquarks. Within this framework, we investigate the projected HL-LHC sensitivity to the LNV same-sign dilepton plus jets signature, $pp \to \ell^{\pm}\ell'^{\pm} + \text{jets}$. The model is further constrained by neutrino mass measurements~\cite{Esteban:2024eli}, charged lepton flavor violation~\cite{Babu:2019mfe}, electroweak precision observables~\cite{Babu:2010vp}, and existing leptoquark searches~\cite{ATLAS:2020dsk,CMS:2024bnj}. We evaluate their impact to map out the currently allowed parameter space, before assessing the collider reach for the lepton number violating signature.

The remainder of this paper is organized as follows. In Sec.~\ref{sec:model}, we present the model framework. Constraints from neutrino mass measurements, low energy measurements, electroweak precision observables, and leptoquark searches at the LHC are discussed in Sec.~\ref{sec:constraints}. In Sec.~\ref{sec:lnv_cs}, we analyze the relevant production channels leading to the LNV same-sign dilepton plus jets signature. In Sec.~\ref{sec:analysis}, we present a detector level collider analysis, where we evaluate the discovery potential at the HL-LHC. We summarize and conclude in Sec.~\ref{sec:conclusion}.

\section{Leptoquark variant of the Zee Model}
\label{sec:model}

The Leptoquark variant of the Zee Model generates small neutrino masses radiatively at the one-loop level~\cite{Zee:1980ai,AristizabalSierra:2007nf}. It preserves the Standard Model (SM) gauge structure $SU(3)_{C} \times SU(2)_L \times U(1)_{Y}$ and extends the SM particle content by introducing a $SU(2)_L$ doublet leptoquark (LQ) $\Omega(3,2,1/6) = (\omega^{2/3}, \omega^{-1/3})$ and a singlet LQ $\chi^{-1/3}(3,1,-1/3)$. The Yukawa Lagrangian for the LQs can be written as
\begin{equation}
    \mathcal{L}_\text{Yuk} \supset \lambda_{\alpha \beta} L_{\alpha}^{i}d_{\beta}^c \Omega^j \epsilon_{ij} + \lambda^\prime_{\alpha \beta} L_{\alpha}^{i} Q_{\beta}^{j} \chi^{*}\epsilon_{ij} + \textrm{h.c.}\,,
    \label{eqn:Yukawa_lagrangian_1}
\end{equation}
where $\{\alpha, \beta\}$ denote the generation indices and $\{i,j\}$ are the $SU(2)_L$ indices. The couplings $\lambda$ and $\lambda^\prime$ are the Yukawa coupling constants for the $\Omega$ and $\chi$ fields. $L$ and $Q$ are, respectively, the left-handed lepton and quark doublets, and $d^c$ represent the conjugate of the right-handed down-type fields.

To ensure proton stability, we assume a global baryon number symmetry under which both $\Omega$ and $\chi^{-1/3}$ carry baryon number $-1/3$. In the absence of this symmetry, dangerous baryon number violating operators of the type $Q Q \chi$ and $u^c d^c \chi^*$ would be allowed, leading to rapid proton decay. This model can be viewed as a special case of $R$-parity violating supersymmetric model~\cite{Dercks:2017lfq}, where the fields $\Omega$ and $\chi^{-1/3}$ are identified as the superpartners of the quark doublet and $d^c$ fields respectively (\textit{i.e.}, $\Omega \sim \tilde{Q}$ and $\chi^{* 1/3} \sim \tilde{d^c}$). 

The Higgs potential contains nontrivial interaction terms given by 
\begin{equation}
V \supset \mu \left(\Omega^\dagger H \chi+ (\Omega^\dagger H \chi)^*\right) + \lambda_7 (H^\dagger \Omega) (\Omega^\dagger H)~. 
\label{eqn:cubic_scalar_coupling}
\end{equation}
Here, we have made the complex parameter $\mu$ real by a field redefinition. This term in the Higgs potential generates mixing between the $\omega^{-1/3}$ and $\chi^{-1/3}$ fields after electroweak symmetry breaking, with the mass-squared matrix given by
\begin{eqnarray}
\mathcal{M}^2_{\omega^{-1/3}-\chi^{-1/3}} = \left(
\begin{matrix}
M_{\omega^{2/3}}^2 + \frac{\lambda_7}{2} v^2 & \sqrt{2} \mu v \\
\sqrt{2} \mu v &  m_\chi^2 
\end{matrix}\right)~,
\end{eqnarray}
where $v=246$~GeV is the vacuum expectation value of the neutral component of the SM Higgs doublet $H$, $M_{\omega^{2/3}}^2$ is the physical mass-squared of $\omega^{2/3}$, which includes a contribution from a $\lambda_4 (H^\dagger H) (\Omega^\dagger \Omega)$ term, and $m_\chi^2$ is the bare mass term plus the contribution from the $\lambda_5 (H^\dagger H)(\chi^\star \chi)$ term. The mixing between $\omega^{-1/3}$ and $\chi^{-1/3}$ results in the mass eigenstates $\xone$ and $\xtwo$,
\begin{align}
    \xone &= \cos\theta_{\rm LQ}\,\omega^{-1/3} + \sin\theta_{\rm LQ}\,\chi^{-1/3}\,,\\
    \xtwo &= -\sin\theta_{\rm LQ}\,\omega^{-1/3} + \cos\theta_{\rm LQ}\,\chi^{-1/3}\,.
    \label{eqn:LQ_mass_eigen}
\end{align}
The mixing angle $\theta_{\rm LQ}$ is given by $\tan 2\theta_{\rm LQ} = 2\sqrt{2}\mu v/(m_{\omega}^2 + \lambda_7 v^2/2 - m_{\chi}^2)$. The squared-masses of the physical eigenstates $\xone$ and $\xtwo$ are given by 
\begin{equation}
    M_{X_{1,2}}^2 = \frac{1}{2} \left[ M_{\om}^2 + \frac{\lambda_7}{2} v^2 + m_{\chi}^2 \mp \sqrt{(M_{\om}^2 + \frac{\lambda_7}{2} v^2 - m_{\chi}^2)^2 + 8\mu^2 v^2}\right].
\end{equation}

The Yukawa Lagrangian in Eq.~\eqref{eqn:Yukawa_lagrangian_1} can be expressed in terms of the physical mass eigenstates as
\begin{eqnarray}
    \mathcal{L}_\text{Yuk}     & = & -\lambda_{\alpha\beta} \ell_{\alpha} d_{\beta}^c \omega^{2/3} - \lambda^\prime_{\alpha\beta} V^\dagger_{\rm CKM} \sin{\theta_{\rm LQ}}\ell_{\alpha} u_\beta \xone - \lambda^\prime_{\alpha\beta} V^\dagger_{\rm CKM} \cos{\theta_{\rm LQ}}\ell_{\alpha}u_{\beta} \xtwo \nonumber \\ 
    && + \left(\lambda_{\alpha\beta}\cos{\theta_{\rm LQ}}\nu_{\alpha}d_{\beta}^c + \lambda^\prime_{\alpha\beta} \sin{\theta_{\rm LQ}} \nu_{\alpha} d_{\beta}\right)\xone \nonumber \\
    && -  \left(\lambda_{\alpha\beta} \sin{\theta_{\rm LQ}} \nu_{\alpha}d_{\beta}^c - \lambda_{\alpha\beta}^\prime \cos{\theta_{\rm LQ}} \nu_{\alpha} d_{\beta}\right)\xtwo + \textrm{h.c.}\,,
    \label{eqn:Yuk_lag_LQ}
\end{eqnarray}
where $\ell$ and $\nu$ are the charged and neutral-lepton components of the lepton doublet, and $u$ is the left-handed up-type quark. 
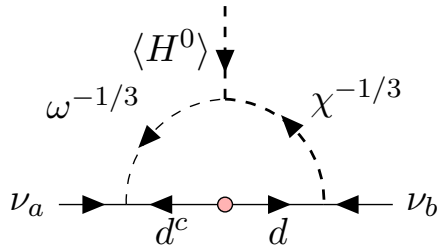
\begin{figure}[!t]
    \centering\scalebox{1.3}{
    \feynmandiagram[horizontal=ii tp f1] {
    ii [particle=\(\nu_{a}\)] -- [fermion] a -- [anti fermion, edge label'=\(d^{c}\)] b [dot] -- [fermion, edge label'=\(d\)] c -- [anti fermion] f1 [particle=\(\nu_{b}\)],
    a -- [anti charged scalar, quarter left, edge label=\(\omega^{-1/3}\)] b1 [above=2cm of b] -- [anti charged scalar, quarter left, edge label=\(\chi^{-1/3}\), dashed, thick] c,
    b1 -- [anti charged scalar, edge label=\(\langle H^0 \rangle\), dashed, thick] d1 [above=4cm of b],
    };}
    \caption{Feynman diagram representing neutrino mass generation at one-loop in the Leptoquark variant of the Zee Model.}
    \label{fig:nu_mass_generation}
\end{figure}
The Yukawa terms in Eq.~\eqref{eqn:Yuk_lag_LQ}, together with the LQ mixing induced by the cubic scalar term in Eq.~\eqref{eqn:cubic_scalar_coupling}, lead to LNV by two units, $|\Delta L| = 2$. This allows neutrino mass generation at one-loop as illustrated with the Feynman diagram in Fig.~\ref{fig:nu_mass_generation}. 

\section{Constraints}
\label{sec:constraints}

The considered model is sensitive to a range of experimental measurements, including neutrino oscillation data, low energy observables, precision electroweak measurements, and leptoquark searches at the LHC. In this section, we examine the impact of these constraints on the model parameter space and identify the currently allowed region, which will serve as the foundation for our study of LNV signatures at the HL-LHC.

\subsection{Neutrino mass constraints}
\label{sec:neutrino_mass_constraints}
Neutrino oscillation data place strong constraints on the mass-squared differences~\cite{Esteban:2024eli}
\begin{equation}
    \Delta m_{21}^2 = 7.49^{+0.19}_{-0.19} \times 10^{-5}~\mathrm{eV}^{2}, \quad \Delta m_{32}^2 = 2.513^{+0.021}_{-0.019} \times 10^{-3}~\mathrm{eV}^2 \,.
    \label{eqn:neutrino_mass_sq_diff}
\end{equation}
In addition, kinematic measurements of Tritium beta decay, ${}^3\mathrm{H} \to {}^3\mathrm{He} + e^- + \bar{\nu}_e$, at the KATRIN experiment constrain the effective electron neutrino mass to $m_{\nu_e} < 0.45~\mathrm{eV}$ at $90\%$ CL~\cite{KATRIN:2024cdt}.\footnote{The parameter $m_{\nu_e}$ is defined as $m_{\nu_e}^2\equiv \sum_i m_i^2 |U_{ei}|^2$, where $U$ is the $3\times 3$ unitary leptonic (PMNS) mixing matrix.} Recent results from NuFit collaboration~\cite{Esteban:2024eli}, performing a global analysis of the neutrino data, imply that for normal ordering the sum of neutrino masses must lie within
\begin{equation}
    0.058~\mathrm{eV} \lesssim \sum m_i \lesssim 1.2~\mathrm{eV}\,\quad \mathrm{at~95\%~CL}.
    \label{eq:mnu-upper}
\end{equation}
In this work, we adopt normal ordering and fix the lightest neutrino mass to $m_1 = 0.04~$eV. The remaining masses are then determined from the measured mass squared differences: $m_2 = \sqrt{m_1^2 + \Delta m_{21}^2}$ and $m_3 = \sqrt{m_2^2 + \Delta m_{32}^2}$. Using the values in Eq.~\eqref{eqn:neutrino_mass_sq_diff}, this yields $\sum m_i \simeq 0.15~\mathrm{eV}$.

In the leptoquark variant of the Zee Model, the neutrino mass matrix in the flavor basis, $M_{\nu}$, arises radiatively at one-loop and takes the form
\begin{equation}
    M_{\nu} = U^\star\,\, \mathrm{diag}(m_1, m_2, m_3)\,\, U^\dagger = \kappa \left(\lambda M_d \lambda^{\prime T} + \lambda^\prime M_d \lambda^T\right)\,,
    \label{eqn:solve_for_yukawa}
\end{equation}
where  the loop-induced coefficient is
\begin{equation}
    \kappa = \left(3/32 \pi^2\right) \sin{2\theta_{\rm LQ}} \ln\left(m_{\xone}^{2}/m_{X_2}^{2}\right)\,.
\end{equation}
 Here, $M_{d}$ is the down-type quark mass matrix and $m_{\xone}$, $m_{X_2}$ denote the masses of the leptoquark fields $\xone$ and $\xtwo$, respectively. We adopt the standard parametrization for the $3\times 3$ PMNS matrix, $U$, from Ref.~\cite{ParticleDataGroup:2024cfk} and use the matrix element values from the NuFit-6.0 global analysis~\cite{Esteban:2024eli}. For simplicity, we assume that the Yukawa matrices are related by a constant scaling factor $c$
 \begin{equation}
     \lambda_{ij} = c\, \lambda_{ij}^{\prime}\,,
 \end{equation}
which reduces Eq.~\eqref{eqn:solve_for_yukawa} to a relation involving a single unknown matrix. In our analysis, for each choice of leptoquark masses and mixing angle, we solve this equation to determine the elements of the $3\times 3$ Yukawa matrices.

\subsection{Low energy measurements}

Various low energy measurements, including rare decays of charged leptons and mesons, place relevant constraints on the parameter space of interest. For instance, leptoquarks can induce charged lepton flavor violating (cLFV) decays of the form $\ell_{\alpha} \to \ell_{\beta} + \gamma$ at one-loop, where the $\gamma$ is radiated either from the leptoquark or the quark lines in the loop. In this class of processes, the contributions mediated by the $\omega$ leptoquark are subject to a GIM-like cancellation~\cite{Lavoura:2003xp,Babu:2010vp,Babu:2019mfe,Crivellin:2020mjs}, resulting in relatively weak constraints on the $\lambda$ couplings. The surviving terms in the amplitude are suppressed by a factor of order $(m_d^2/m_\omega^2)$. In contrast, the processes mediated by $\xone$ and $\xtwo$, owing to their $\chi^{-1/3}$ admixture, can introduce strong constraints on the $\lambda^\prime$ couplings. The branching ratio for the $\ell_\alpha \to \ell_\beta + \gamma$ process mediated by $\xone$ and $\xtwo$ is given by 
\begin{equation}
    \mathcal{BR}(\ell_\alpha \to \ell_\beta + \gamma) = \frac{9\alpha_{\mathrm{em}}\,m_{\ell_\alpha}^5}{576\,(16\pi^2)^2 \, \Gamma(\ell_\alpha)} |\lambda_{\beta d}^\prime \lambda_{\alpha d}^{\prime\star}|^2 \left(\frac{\sin^2\theta_{\rm LQ}}{m_{\xone}^2} + \frac{\cos^2\theta_{\rm LQ}}{m_{\xtwo}^2}\right)^2.
\end{equation}
In Table~\ref{tab:constraints_new}, we present the resulting constraints on $|\lambda_{\beta d}^\prime \lambda_{\alpha d}^{\prime\star}|$ for $\mu \to e \gamma$, $\tau\to e\gamma$, and $\tau \to \mu \gamma$, as a function of $\theta_{\rm LQ}$ and leptoquark masses $m_{\xone}$ and $m_{\xtwo}$. 

\begin{table}[!t]
    \centering\scalebox{0.72}{
    \begin{tabular}{|c|c|c|} \hline  
       Process & Current Limit & Bounds~$\left(\mathrm{TeV}\right)^{-2}$  \\ \hline \hline
       $\mu \to e \gamma$ & $ \mathcal{BR} < 1.5 \times 10^{-13}$~\cite{MEGII:2025gzr} & $|\lambda^{\prime}_{ed}\lambda^{\prime *}_{\mu d}| < 8.6 \times 10^{-4}\left(\frac{s_\theta^2}{m_{\xone}^2} + \frac{c_{\theta}^2}{m_{\xtwo}^2}\right)^{-1}$ \\
       $\tau \to e \gamma$ & $\mathcal{BR} < 3.3 \times 10^{-8}$~\cite{BaBar:2009hkt} & $|\lambda^{\prime}_{ed}\lambda^{\prime *}_{\tau d}| < 1.6\left(\frac{s_\theta^2}{m_{\xone}^2} + \frac{c_{\theta}^2}{m_{\xtwo}^2}\right)^{-1}$ \\
       $\tau \to \mu \gamma$ &  $\mathcal{BR} < 4.2 \times 10^{-8}$~\cite{Belle:2021ysv} & $|\lambda^{\prime *}_{\mu d}\lambda^{\prime}_{\tau d}| < 1.8 \left(\frac{s_\theta^2}{m_{\xone}^2} + \frac{c_{\theta}^2}{m_{\xtwo}^2}\right)^{-1}$ \\ \hline 
       \multicolumn{3}{|l|}{$\mu N \to e N$}\\ \cdashline{1-3}
       ${}^{48}_{22}\mathrm{Ti}$ & $\mathcal{BR} < 6.1 \times 10^{-13}$ & $\frac{|\lambda^\star_{ed}\lambda_{\mu d}|}{m_{\om}^2} + |\lambda^{\star\prime}_{ed}\lambda^\prime_{\mu d}| \left(\frac{s^2_\theta}{m_{\xone}^2} + \frac{c^2_\theta}{m_{\xtwo}^2}\right) < 4.30 \times 10^{-6}$ \\ 
       ${}^{197}_{79}\mathrm{Au}$ & $\mathcal{BR} < 7.0 \times 10^{-13}$~\cite{Bertl:55087} & $\frac{|\lambda^\star_{ed}\lambda_{\mu d}|}{m_{\om}^2} + |\lambda^{\star\prime}_{ed}\lambda^\prime_{\mu d}| \left(\frac{s^2_\theta}{m_{\xone}^2} + \frac{c^2_\theta}{m_{\xtwo}^2}\right) < 4.29 \times 10^{-6}$ \\ 
       ${}^{208}_{82}\mathrm{Pb}$ & $\mathcal{BR} < 4.6 \times 10^{-11}$~\cite{PhysRevLett.76.200} & $\frac{|\lambda^\star_{ed}\lambda_{\mu d}|}{m_{\om}^2} + |\lambda^{\star\prime}_{ed}\lambda^\prime_{\mu d}| \left(\frac{s^2_\theta}{m_{\xone}^2} + \frac{c^2_\theta}{m_{\xtwo}^2}\right) < 3.56 \times 10^{-5}$ \\ \hline 
        $K_L^0 \to e^+ e^-$ & $\mathcal{BR} < 8.7^{+5.7}_{-4.1} \times 10^{-12}$~\cite{BNLE871:1998bii} & $|\lambda_{ed}^{*} \lambda_{e s}| \left(\frac{1}{m_{\omega^{2/3}}^{2}} \right) < 1.9 \times  10^{-3}$ \\
        $K_L^0 \to \mu^+ \mu^-$ & $\mathcal{BR} < \left( 6.84 \pm 0.11\right) \times 10^{-9}$~\cite{ParticleDataGroup:2026aaa} & $|\lambda_{\mu d}^{*} \lambda_{\mu s}|\left(\frac{1}{m_{\omega^{2/3}}^{2}} \right) < 4.7 \times  10^{-5}$\\
        $K_L^0 \to e^\pm \mu^\mp$ & $\mathcal{BR} < 4.7 \times 10^{-12}$~\cite{BNL:1998apv} &  $|\lambda_{es} \lambda_{\mu d}^{\star} + \lambda_{ed} \lambda_{\mu s}^{\star}|\left(\frac{1}{m_{\omega^{2/3}}^{2}} \right)  < 1.9 \times  10^{-5}$ \\
        $K^+ \to \pi^+ e^+ e^-$ & $\mathcal{BR} < \left(3.00 \pm 0.09 \right) \times 10^{-7}$~\cite{ParticleDataGroup:2026aaa} & $|\lambda_{ed}^{*} \lambda_{es}|\left(\frac{1}{m_{\omega^{2/3}}^{2}} \right) < 2.3 \times  10^{-2}$ \\
        $K^+ \to \pi^+ \mu^+ \mu^-$ & $\mathcal{BR} < \left(9.15 \pm 0.08 \right) \times 10^{-8}$~\cite{NA62:2022qes} & $|\lambda_{\mu d}^{*} \lambda_{\mu s}|\left(\frac{1}{m_{\omega^{2/3}}^{2}} \right) < 1.85 \times  10^{-2}$ \\
        $K^+ \to \pi^+ \mu^+ e^-$ & $\mathcal{BR} < 1.3 \times 10^{-11}$~\cite{Sher:2005sp} & $|\lambda_{e d}^{*} \lambda_{\mu s}|\left(\frac{1}{m_{\omega^{2/3}}^{2}} \right),\,|\lambda_{e s} \lambda_{\mu d}|^{*}\left(\frac{1}{m_{\omega^{2/3}}^{2}} \right) < 1.9 \times  10^{-4}$ \\
        $K^+ \to \pi^+ \nu \bar{\nu}$ & $\mathcal{BR} < 9.6^{+1.9}_{-1.8} \times 10^{-11}$~\cite{NA62:2026rwr} & $\sum\limits_{m\neq n} \left| \left(\frac{s_\theta^2}{m_{X_1}^2} + \frac{c_\theta^2}{m_{X_2}^2} \right) \hat{\lambda}_{dn}^\prime \hat{\lambda}_{sm}^\prime - \left(\frac{c_\theta^2}{m_{X_1}^2} + \frac{s_\theta^2}{m_{X_2}^2} \right) \hat{\lambda}_{sn}^\prime \hat{\lambda}_{dm}^\prime \right| < 3.1 \times 10^{-4}$ \\ \hline 
    \end{tabular}}
    \caption{Summary of constraints from low energy measurements on the leptoquark Yukawa couplings $\lambda$ and $\lambda^\prime$ as a function of leptoquark masses $\{m_{\om}, m_{\xone}, m_{\xtwo}\}$~\cite{PhysRevLett.76.200,BNLE871:1998bii,BNL:1998apv,Sher:2005sp,Bertl:55087,BaBar:2009hkt,Babu:2010vp,Babu:2019mfe,Mandal:2019gff,Babu:2020hun, Belle:2021ysv,NA62:2022qes,MEGII:2025gzr,ParticleDataGroup:2026aaa,NA62:2026rwr}. In the expression for the $K^+ \to \pi^+ \nu \bar{\nu}$ constraint, $\hat{\lambda}$ and $\hat{\lambda}^\prime$ are defined as $\hat{\lambda} = \lambda^T U$ and $\hat{\lambda}^\prime = \lambda^{\prime T} U$~\cite{Mandal:2019gff}, where $U$ is the PMNS matrix~(see Sec.~\ref{sec:neutrino_mass_constraints} for details). The indices $m,n = 1,2,3$ appearing in the last line denote the neutrino states in the mass basis, following the convention of \cite{Mandal:2019gff}.}
    \label{tab:constraints_new}
\end{table}

Another rare cLFV process that probes both $\lambda$ and $\lambda^\prime$ couplings is coherent $\mu$–$e$ conversion in nuclei, $\mu N \to e N$, which occurs at tree level in these models. The corresponding branching ratio can be expressed as~\cite{Babu:2010vp,Babu:2019mfe}
\begin{eqnarray}
    \mathcal{BR}(\mu N \to e N) &=& \frac{|\vec{p_e}\,E_e\,m_{\mu}^{3}\,G_F^2\,\alpha_{\mathrm{em}}^3\,Z_{eff}^4\,F_p^2|}{64\,\pi^2\,Z\,\Gamma_N} \left(2 A - Z\right)^2 \nonumber  \\  
    &\times& \left(\frac{|\lambda^\star_{ed}\lambda_{\mu d}|}{m_{\om}^2} + |\lambda^{\star\prime}_{ed}\lambda^\prime_{\mu d}| \left(\frac{\sin^2\theta_{\rm LQ}}{m_{\xone}^2} + \frac{\cos^2\theta_{\rm LQ}}{m_{\xtwo}^2}\right)\right)^2\,,
\end{eqnarray}
where $\vec{p}_e$ and $E_e$ denote the momentum and energy of the outgoing electron, $Z_{eff}$ is the effective atomic number, $F_p$ is the nuclear matrix element, $A$ and $Z$ are the mass number and atomic number of the nucleus, respectively. $\Gamma_N$ is the rate of capture of muons by the nucleus $N$. Using the nuclear and model input parameters detailed in Ref.~\cite{Babu:2019mfe}, we derive upper limits on the Yukawa couplings from current measurements of $\mathcal{BR}(\mu N \to e N)$ for ${}^{48}_{22}\mathrm{Ti}$, ${}^{197}_{79}\mathrm{Au}$ and ${}^{208}_{82}\mathrm{Pb}$, as summarized in~Table~\ref{tab:constraints_new}. Rare kaon decays provide additional probes of the flavour structure of leptoquark interactions, constraining the product of leptoquark Yukawa couplings involving the first and second generation down-type quarks. The adopted limits are summarized in Table~\ref{tab:constraints_new}. We find that $\mu \to e\gamma$ and $\mu N \to e N$ processes are the most stringent among the cLFV and meson decay constraints on our parameter space. A detailed overview of these processes and their impact on the LQ-variant of the Zee Model can be found in Refs.~\cite{Babu:2010vp,Babu:2019mfe,Babu:2020hun}. 

\subsection{Electroweak precision constraints}

Precision electroweak observables provide additional restrictions via the oblique parameters $S$, $T$, and $U$~\cite{Grimus:2008nb,Babu:2019mfe,Crivellin:2020ukd}. In our model, they take the form
\begin{align}
    S =& \frac{1}{12\pi} \left( \sin^2\theta_{\rm LQ} \,\ln\frac{m_{\xone}^2}{m_{\om}^2} + \cos^2\theta_{\rm LQ} \,\ln\frac{m_{\xtwo}^2}{m_{\om}^2} + 3 \cos^{2}\theta_{\rm LQ}\sin^2\theta_{\rm LQ}\, \mathcal{G}\left(\frac{m_{\xone}^2}{m_{\xtwo}^2}\right)\right) ,  \\
    T =& \frac{3}{8 \pi^2 \alpha_{\textrm{em}}v^2} \Bigg(\cos^2{\theta_{\rm LQ}} \mathcal{F}(m_{\om}^2, m_{\xone}^2)\, +\, \sin^2{\theta_{\rm LQ}} \mathcal{F}(m_{\om}^2, m_{\xtwo}^2)\, \nonumber \\
    &- \,\cos^2\theta_{\rm LQ} \sin^2{\theta_{\rm LQ}} \mathcal{F}(m_{\xone}^2, m_{\xtwo}^2) \Bigg ), \\
    U =& \frac{1}{4\pi} \left( \sin^2\theta_{\rm LQ} \,\mathcal{G}\left(\frac{m_{\om}^2}{m_{\xone}^2}\right) + \cos^2\theta_{\rm LQ} \,\mathcal{G}\left(\frac{m_{\om}^2}{m_{\xtwo}^2}\right) - \cos^{2}\theta_{\rm LQ}\sin^2\theta_{\rm LQ}\, \mathcal{G}\left(\frac{m_{\xone}^2}{m_{\xtwo}^2}\right)\right) ,   
\end{align}
with 
\begin{eqnarray}
    &\mathcal{F}(a,b)& = \frac{1}{2}(a + b) - \frac{ab}{a - b} \ln\left(\frac{a}{b}\right),  \\
    &\mathcal{G}(c)& = \frac{c^3 - 3 c^2 - 3c + 1}{(c-1)^3} \ln\,c - \frac{5c^2 - 22c + 5}{3(c-1)^2}.
\end{eqnarray}

The current data yield the following $2\sigma$ ranges~\cite{ParticleDataGroup:2024cfk}
\begin{equation}
-0.24 < \Delta S < 0.26, ~~ -0.23 < \Delta T < 0.25,~~\mathrm{and}~~-0.19 < \Delta U < 0.17\,,
\label{eqn:oblique_constraints}
\end{equation}
which we impose in our parameter scans. While $\Delta T$ is controlled by
isospin-breaking effects and therefore typically favors small mass
splittings among the leptoquark mass eigenstates, $\Delta S$ and $\Delta U$ depend logarithmically on the leptoquark mass ratios and vanish in the
degenerate mass limit. In practice, the most stringent constraint among the three generally arises from $\Delta T$, as we demonstrate explicitly in Sec.~\ref{sec:combined_constraints}.

\subsection{LHC bounds}

\begin{figure}[!t]
    \centering
    \includegraphics[width=0.6\linewidth]{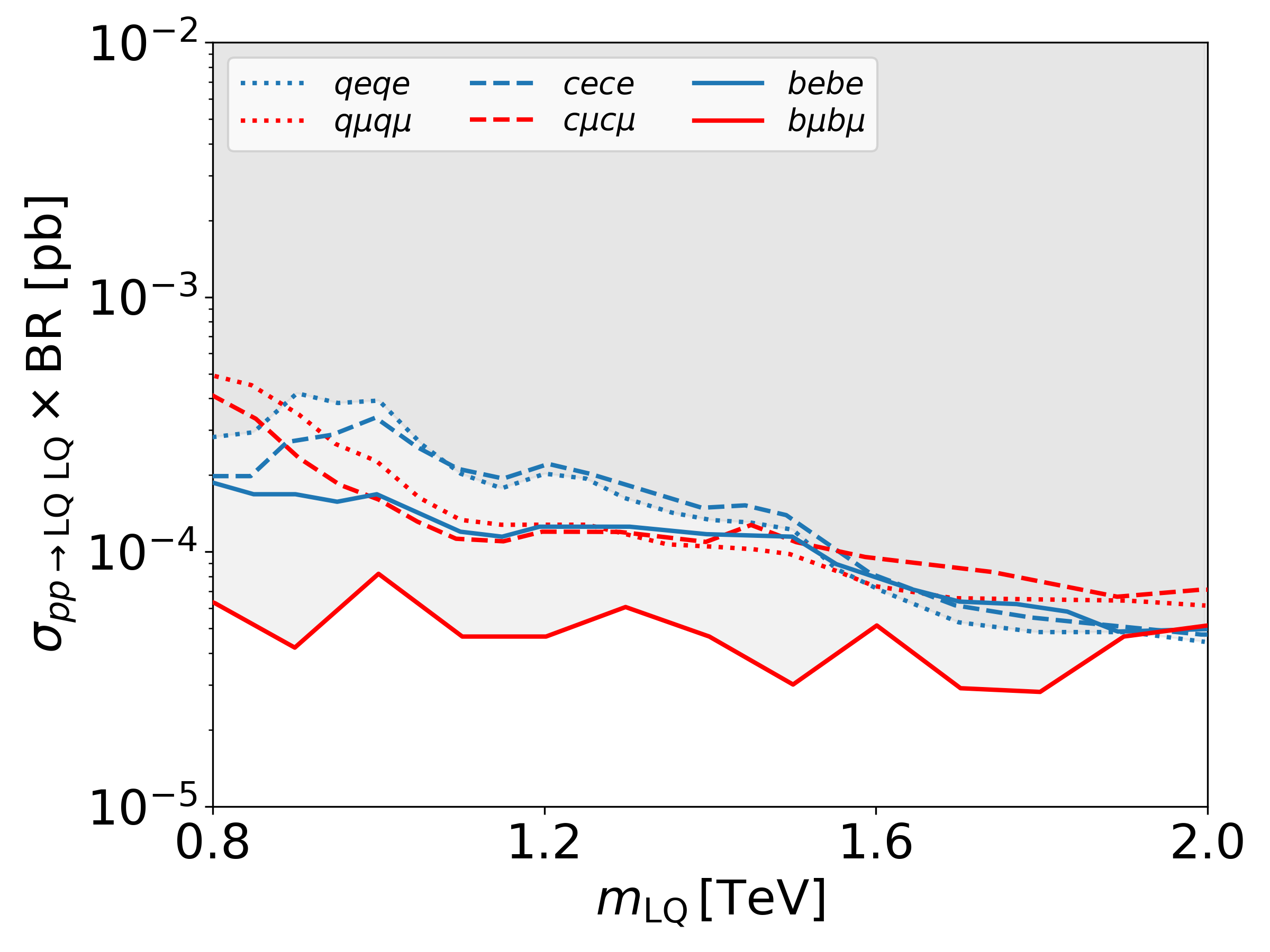}
    \caption{Upper limits on the leptoquark pair production cross-section times branching fraction, $pp \to \mathrm{LQ}\,\overline{{\rm LQ}}\to X$, as a function of leptoquark mass $m_{\mathrm{LQ}}$, from searches in the $X = \{qe qe,\, q\mu q\mu,\, cece,\, c\mu c\mu,\, be be,\, b\mu b\mu\}$ channels by ATLAS~\cite{ATLAS:2020dsk} and CMS~\cite{CMS:2024bnj}  at the $\sqrt{s}=13~\text{TeV}$ LHC.}
\label{fig:lhc_LQbr_limits}
\end{figure}

Leptoquark searches at the LHC have been performed in both single and pair production channels~\cite{ATLAS:2024huc, ATLAS:2024lpr, ATLAS:2023uox, ATLAS:2023vxj, ATLAS:2023prb, ATLAS:2020dsk, Varzielas:2023qlb}. The pair production mode is dominated by gluon-gluon and quark-antiquark annihilation, with a cross-section that is nearly independent of the leptoquark Yukawa couplings and depends primarily on the leptoquark mass. The ATLAS and CMS collaborations have derived upper limits on the pair production cross-section multiplied by the branching fraction as a function of the leptoquark mass, which can be directly interpreted in the parameter space relevant to our analysis. In Fig.~\ref{fig:lhc_LQbr_limits}, we show the corresponding upper limits from recent searches at the $\sqrt{s}=13~$TeV LHC using $\mathcal{L}\sim 139~\mathrm{fb}^{-1}$ and $138~\mathrm{fb}^{-1}$ of data collected by ATLAS and CMS, respectively~\cite{CMS:2024bnj, ATLAS:2020dsk}. While the observation of one or more of these signals would provide unambiguous evidence of physics beyond the Standard Model, it would not indicate LNV, as events in these searches are selected by requiring an oppositely charged electron or muon, resulting in final states with zero net lepton number. 

\subsection{Combined constraints}
\label{sec:combined_constraints}

\begin{figure}[!t]
    \centering
    \includegraphics[width=0.65\linewidth]{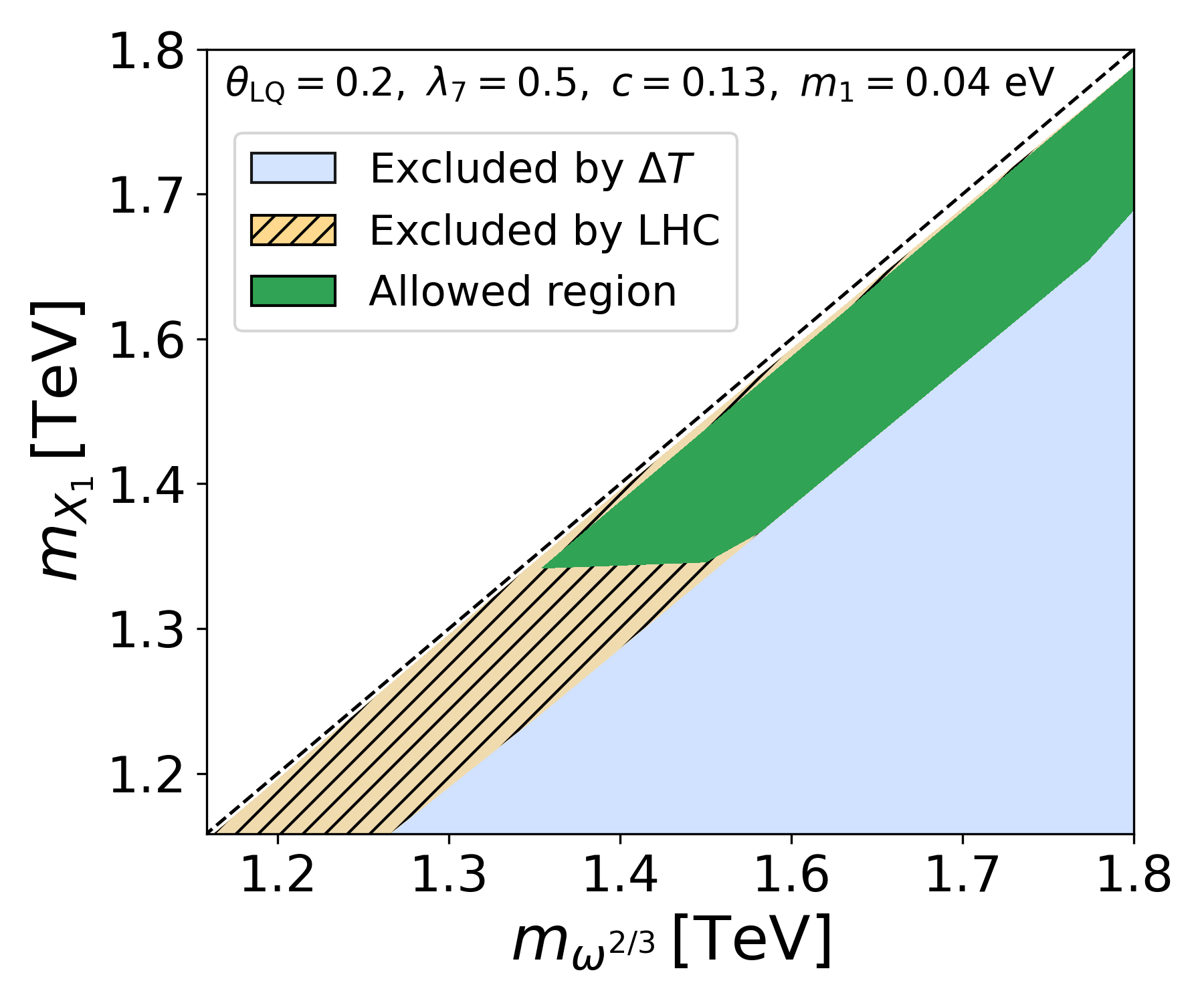}
    \caption{Parameter space is shown in the $m_{\omega^{2/3}}$-$m_{\xone}$ plane. The blue shaded region is excluded by constraints from $\Delta T$ at 95\%~CL, while the yellow shaded region is ruled out by leptoquark pair production searches at the $\sqrt{s}=13~$TeV LHC~\cite{ATLAS:2020dsk,CMS:2024bnj}. The green shaded region denotes the parameter space consistent with current constraints. Results are shown for the benchmark scenario $\{\theta_\mathrm{LQ} = 0.2,\ \lambda_7 = 0.5,\, c \equiv \lambda_{ij}/\lambda_{ij}^\prime = 0.13, \ m_1 = 0.04~\mathrm{eV}\}$.}
    \label{fig:allowed_param_space}
\end{figure}

In Fig.~\ref{fig:allowed_param_space}, we show the region of parameter space in the $m_{\omega^{2/3}}$–$m_{\xone}$ plane that is consistent with current constraints. Among the three oblique parameters, $\Delta T$ imposes the most stringent constraint, allowing only a narrow band of small mass splittings, $\Delta m=m_{\omega^{2/3}}-m_{\xone}$. For illustration, the region excluded by the 95\% CL constraint on $\Delta T$ is displayed as a blue-shaded area in Fig.~\ref{fig:allowed_param_space} for a representative benchmark point with $\theta_\mathrm{LQ} = 0.2$, $\lambda_7 = 0.5$, $c \equiv \lambda_{ij}/\lambda_{ij}^\prime = 0.13$, and $m_1 = 0.04~\mathrm{eV}$.  By contrast, the constraints from $\Delta S$ and $\Delta U$, as well as those from low energy observables, are significantly weaker and do not restrict the scanned parameter region. The stronger sensitivity of $\Delta T$ arises because it is particularly sensitive to isospin-breaking effects, {\it i.e.}, mass splittings within weak multiplets, whereas $\Delta S$ and $\Delta U$ are less affected by such splittings.

The region allowed by $\Delta T$ is further constrained by leptoquark searches at the LHC. Existing upper limits on the leptoquark pair production cross-section times branching fraction exclude leptoquark masses $m_{\rm LQ}\lesssim 1.4~$TeV. The parameter region consistent with all the current constraints, including neutrino mass bounds, low energy measurements, oblique parameters, and LHC limits, is illustrated as the green shaded region in Fig.~\ref{fig:allowed_param_space}. In the following, we focus on this allowed region, characterized by small mass splittings, $\Delta m=m_{\omega^{2/3}}-m_{\xone}$, and leptoquark masses $m_{\rm LQ}\gtrsim 1.4~$TeV, to explore the potential for observing a signature of lepton number violation at the HL-LHC.

\section{LNV signature at the LHC}
\label{sec:lnv_cs}

\begin{figure}[!t]
\centering
\includegraphics[height=3.2cm, width=4.5cm]{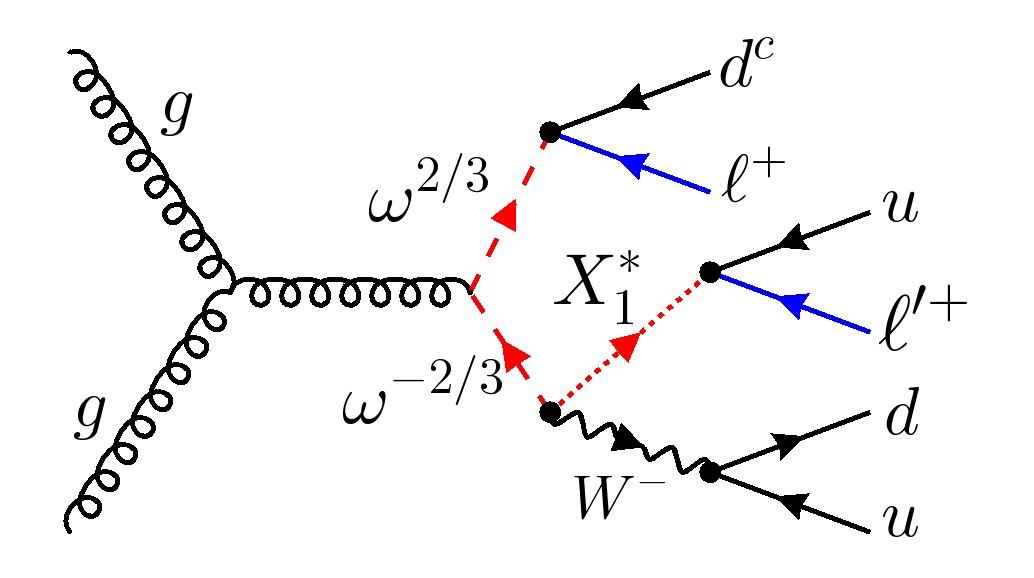} \hspace{0.01cm}
\includegraphics[height=3.2cm, width=4.5cm]{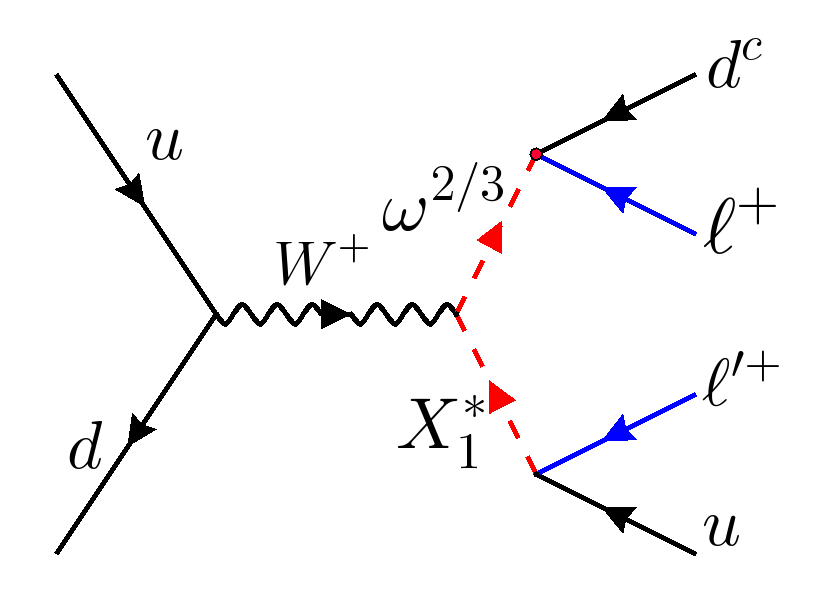} \hspace{0.01cm}
\includegraphics[height=3.2cm, width=4.5cm]{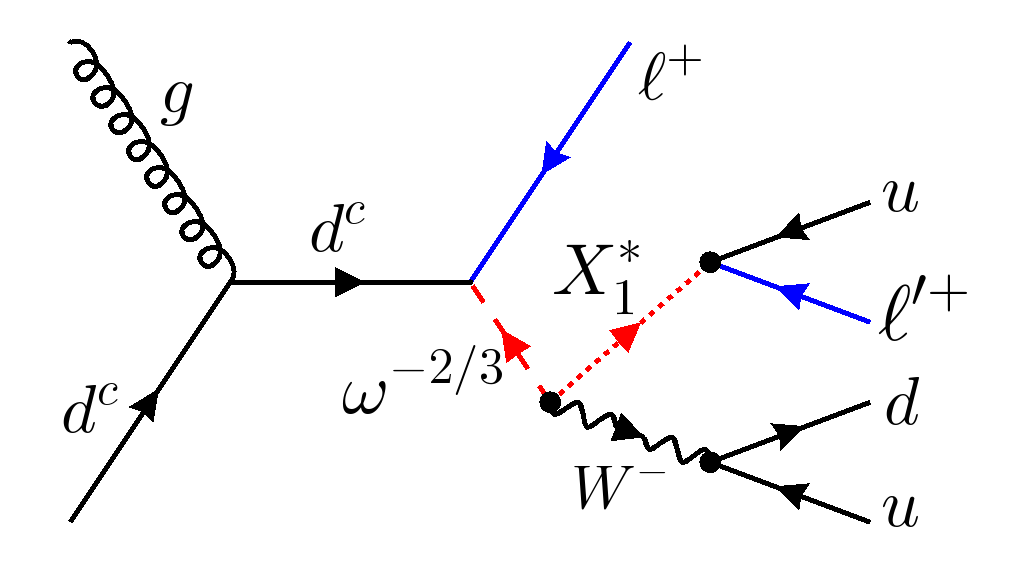}
\caption{Representative Feynman diagrams illustrating the QCD pair production of leptoquarks (left), electroweak pair production~(center), and single leptoquark production (right), leading to LNV signatures $pp \to \ell^{\pm}\ell'^{\pm} +$~jets at the LHC.}
    \label{fig:lnv_feyn}
\end{figure}

The characteristic signature of LNV by two units, $|\Delta L| = 2$, at the LHC is the production of two same-sign charged leptons accompanied by jets, $pp \to \ell^{\pm}\ell'^{\pm} + \text{jets}$. In the parameter space of interest, this final state can arise from cascade decays of both pair and singly-produced leptoquarks. We classify the production mechanisms into three categories: QCD pair production, electroweak (EW) pair production, and single leptoquark production, as illustrated in Fig.~\ref{fig:lnv_feyn}. 

\begin{figure}[!t]
\centering
\includegraphics[height=2.5cm, width=5.0cm]{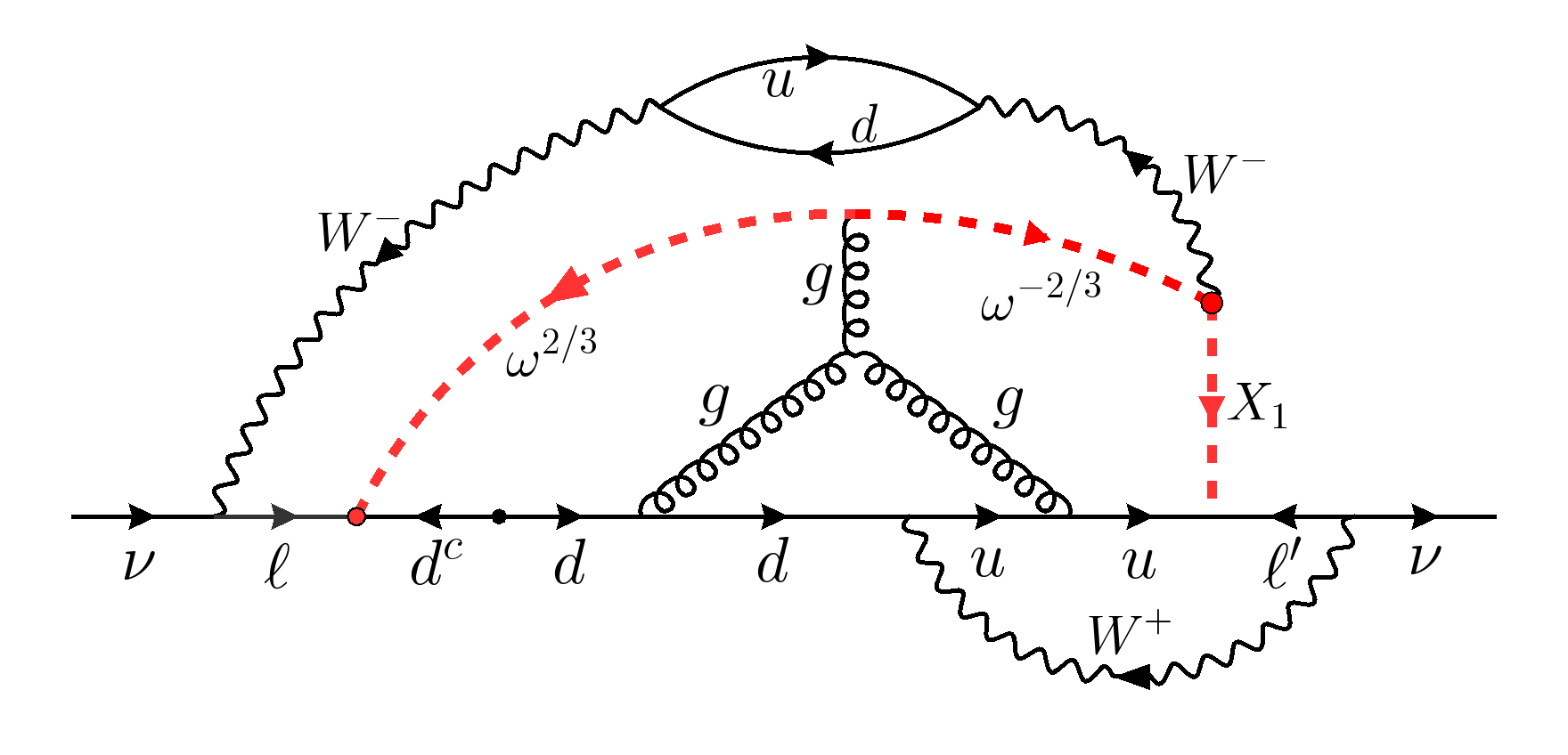}
\includegraphics[height=2.0cm, width=4.2cm]{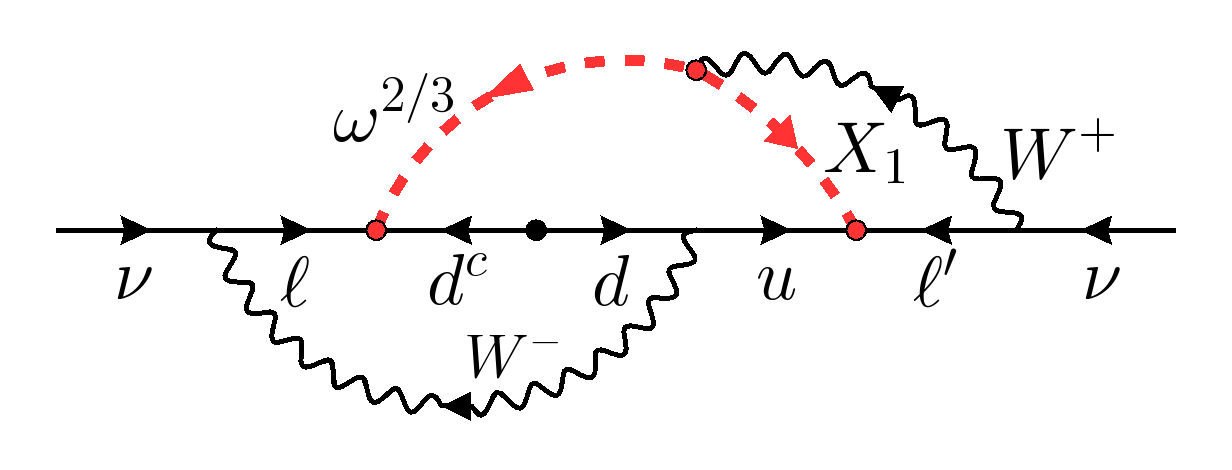}
\includegraphics[height=2.0cm, width=5.0cm]{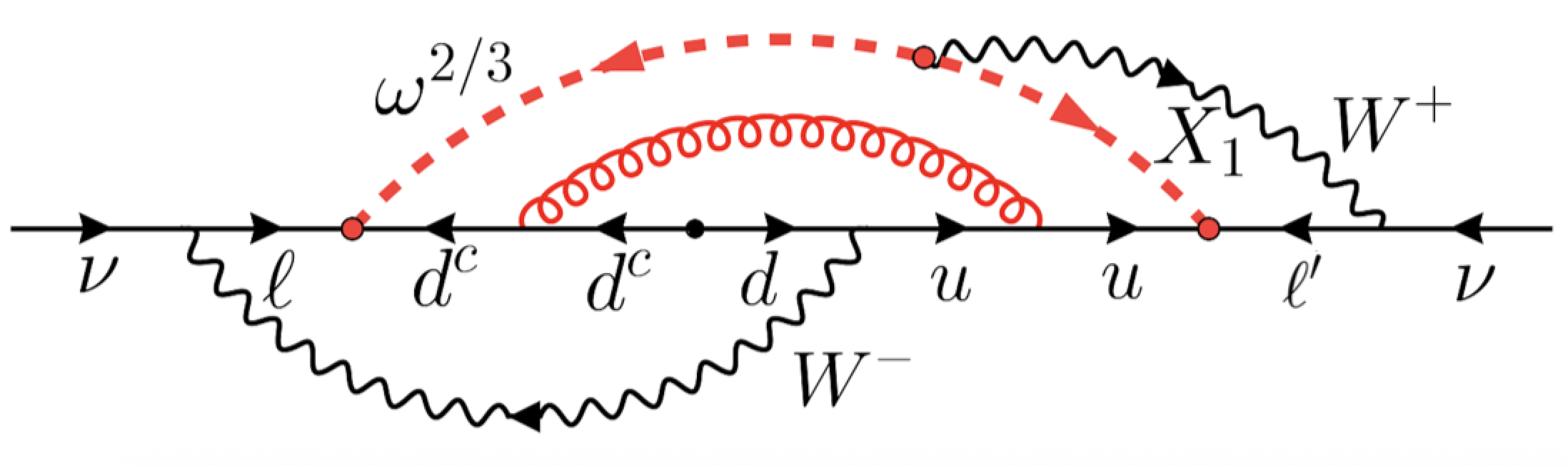}
\caption{Diagrams illustrating the connection between Majorana neutrino mass generation and LNV signatures at the LHC~\cite{Schechter:1981bd,Babu:2022ycv}. The left, center, and right diagrams for neutrino mass generation correspond, respectively, to the LHC processes shown in the left, center, and right panels of Fig.~\ref{fig:lnv_feyn}.}
    \label{fig:lnumass_feyn}
\end{figure}

The theoretical importance of such a $|\Delta L| = 2$ signal lies in its intimate connection to Majorana neutrino mass generation. The generalized form of the  black-box theorem implies that any mechanism producing LNV at colliders necessarily feeds into a Majorana mass term for neutrinos~\cite{Schechter:1981bd,Babu:2022ycv}. In contrast to neutrinoless double beta decay searches, which probe only the electron channel, collider experiments can access a wide range of lepton-flavor combinations, providing complementary and potentially unique insights into the origin of neutrino mass. A schematic illustration of this correspondence between LNV collider signatures and Majorana mass generation is shown in Figs.~\ref{fig:lnv_feyn} and \ref{fig:lnumass_feyn}. 
In the following, we discuss each of the three leading mechanisms for LNV production at the LHC.

\noindent \textbf{1. QCD pair production of leptoquarks:} 
    \begin{eqnarray}
        pp & \to& \om \omstar \to \left(\om \to \ell^{+} j\right)\,\left(\omstar \to (\xonestar \to \ell^+ j)\,(W^- \to j j)\right)\,, \quad \quad \\
        pp &\to& \xtwo\xonestar  \to \left(\xtwo \to (\om \to \ell^+ j)(W^- \to jj)\right)\,\left(\xonestar \to \ell^+ j\right)\,, \quad \quad \\
        pp &\to& \xtwo \xtwostar \to \left(\xtwo \to (\om \to \ell^+ j)(W^- \to jj)\right)\,  \left(\xtwostar \to \ell^+ j\right). \quad \quad 
    \end{eqnarray}
     In this production mode, QCD pair production of leptoquarks is followed by leptoquark decays through different channels to yield the $pp \to \ell^{\pm}\ell'^{\pm} + \text{jets}$ signature. A representative Feynman diagram for this channel is shown in Fig.~\ref{fig:lnv_feyn} (left panel). For instance, in the case of $pp \to \om \omstar$, $\om$ may decay into $\om \to \ell^{+} j$, while its antiparticle $\omstar$ can undergo either on-shell decay $\omstar \to \xonestar W^-$, if $m_{\om} - m_{\xone} \geq m_W$, followed by $\xonestar \to \ell^{+} j$ and $W^- \to jj$, or three-body decay $\omstar \to \xonestar j j$, with $\xonestar \to \ell^{+} j$. In our analysis presented in Sec.~\ref{sec:analysis}, we account for the full matrix element generation for all relevant signal channels, incorporating both on-shell and off-shell contributions. 

\vspace{0.5cm}

\noindent \textbf{2. Electroweak pair production of leptoquarks}
    \begin{equation}
        pp \to \om {X^*_{1,2}} \to \left(\om \to \ell^{+} j\right)\left({X^*_{1,2}} \to \ell^{+} j\right)\,.
    \end{equation}
    This process is generated by quark-quark scattering, $q\bar{q}' \to \om {X^*_{1,2}}$, mediated by s-channel $W$ boson exchange. A representative Feynman diagram is shown in Fig.~\ref{fig:lnv_feyn} (central panel). The relevant vertices $W^{-}\om \xone$ and $W^{-}\om \xtwo$ originate from the Higgs trilinear coupling, with interaction strengths proportional to $\cos \theta_{\rm LQ}$ and $\sin \theta_{\rm LQ}$, respectively. Further model details are provided in Sec.~\ref{sec:model}.

    To realize the LNV signature, both leptoquarks, $\om$ and $X^*_{1,2}$, must decay into ``visible" final states: $\om \to \ell^+ j$ and $X^*_{1,2} \to \ell^+ j$. However, when the mass splitting $m_{\om}-m_{X_{1,2}}$ exceeds the $W$ boson mass, the decay $\om \to W^+ X_{1,2}$ can become dominant. For smaller mass splittings $m_{\om} - m_{X_{1,2}} < m_W$, the three-body decays $\om \to j j \xone$ can still have appreciable rates for small values of $\lambda$.  The visible decay mode $\xone \to \ell^- u$, controlled by $\lambda^\prime \sin\theta_{\rm LQ}$, competes with invisible channels such as $\xone \to \nu d$ and $\xone \to \nu u$, which are sensitive to $\lambda^\prime \sin\theta_{\rm LQ}$ and $\lambda \cos\theta_{\rm LQ}$, respectively. Since neutrino final states obscure the lepton number information, the LNV signal is  appreciable when visible modes dominate, {\it i.e.}, when $\lambda^\prime \sin\theta_{\rm LQ} > \lambda \cos\theta_{\rm LQ}$. The case of $\xtwo$ is qualitatively similar to $\xone$, with the favorable region for the visible decay channel being $\lambda^\prime \cos\theta_{\rm LQ} > \lambda \sin\theta_{\rm LQ}$.

\vspace{0.5cm}

\noindent \textbf{3. Single leptoquark production:}
    \begin{flalign}
        pp \to~ &\omstar \ell^{+} \to \left(\omstar \to (\xonestar \to \ell^+ j)\,(W^- \to jj)\right)\, \ell^{+}\,,  \\
       pp \to~ &\xtwo \ell^{+}  \to \left(\xtwo \to (\om \to \ell^+ j)(W^- \to j j) \right) \ell^{+}\,,  \\
       pp \to~ &{X_{1,2}}^* \ell^+ j \to \left({X_{1,2}}^* \to \ell^+ j \right) \ell^+ j\,,\\
       pp \to~ &\om \ell^+ j \to \left({\om} \to \ell^+ j \right) \ell^+ j\,.  
    \end{flalign}
    These processes involve leptoquark production in association with charged leptons and jets, primarily via quark-quark and quark-gluon fusion. We show a representative Feynman diagram in Fig.~\ref{fig:lnv_feyn} (right panel). 
    
\begin{table}[!t]
    \centering
    \begin{tabular}{|c|c|c|c|} \hline 
       \multicolumn{2}{|c|}{Production modes} & \multicolumn{2}{c|}{LNV signature: $pp \to \ell^{\pm} \ell^{\prime\pm} + \text{jets}$} \\ \hline \hline 
     \multirow{6}{*}{\rotatebox[origin=c]{90}{QCD}} & \multirow{2}{*}{$\om \omstar$} & $\om \to \ell^{+} j$ & $\omstar \to (\xonestar \to \ell^+ j) (W^- \to \text{jets}) $ \\ 
       & & $\omstar \to \ell^{-} j$ & $\om \to (\xone \to \ell^- j) (W^+ \to \text{jets}) $ \\ \cline{2-4}
       & $\xtwo\xonestar$ & $\xonestar \to \ell^+ j$ & $\xtwo \to (\om \to \ell^+ j)(W^- \to \text{jets})$ \\ 
       & $\xone\xtwostar$ & $\xone \to \ell^- j$ & $\xtwostar \to (\omstar \to \ell^- j)(W^+ \to \text{jets})$ \\ \cline{2-4} 
       & \multirow{2}{*}{$\xtwo\xtwostar$} & $\xtwostar \to \ell^+ j$ & $\xtwo \to (\om \to \ell^+ j)(W^- \to \text{jets})$ \\ 
       & &  $\xtwo \to \ell^- j$ & $\xtwostar \to (\omstar \to \ell^- j)(W^+ \to \text{jets})$ \\ \hline \hline 
       \multirow{4}{*}{\rotatebox[origin=c]{90}{EW}} & $\om\xonestar$ & $\om \to \ell^+ j$ & $\xonestar \to \ell^{+} j$ \\
         & $\omstar\xone$ & $\omstar \to \ell^- j$ & $\xone \to \ell^{-} j$ \\ \cline{2-4} 
         & $\om\xtwostar$ & $\om \to \ell^+ j$ & $\xtwostar \to \ell^{+} j$ \\
         & $\omstar\xtwo$ & $\omstar \to \ell^- j$ & $\xtwo \to \ell^{-} j$ \\ \hline \hline 
        \multirow{8}{*}{\rotatebox[origin=c]{90}{Single-LQ}} & $\omstar\ell^+$ & \multicolumn{2}{c|}{$\omstar \to (\xonestar \to \ell^+ j) (W^- \to \text{jets}) $} \\
        & $\om\ell^-$ & \multicolumn{2}{c|}{$\om \to (\xone \to \ell^- j) (W^+ \to \text{jets}) $} \\
        & $\xtwo\ell^+$ & \multicolumn{2}{c|}{$\xtwo \to (\om \to \ell^+ j) (W^- \to \text{jets}) $} \\
        & $\xtwostar\ell^-$ & \multicolumn{2}{c|}{$\xtwostar \to (\omstar \to \ell^- j) (W^+ \to \text{jets}) $} \\ 
        & ${X_{1,2}^*}\ell^+ j $ & \multicolumn{2}{c|}{${X_{1,2}^*} \to \ell^+ j  $} \\
        & $X_{1,2}\ell^- j $ & \multicolumn{2}{c|}{$X_{1,2} \to \ell^- j  $} \\
        & $\om\ell^+ j $ & \multicolumn{2}{c|}{$\om \to \ell^+ j $} \\
        & $\omstar\ell^- j $ & \multicolumn{2}{c|}{$\omstar \to \ell^- j $} \\\hline 
    \end{tabular}
    \caption{Summary of QCD and EW mediated leptoquark pair production, as well as single leptoquark production channels, within the leptoquark extension of the Zee Model, together with their decay topologies that give rise to the LNV, $|\Delta L| = 2$, same-sign dilepton plus jets final state, $pp \to \ell^{\pm} \ell'^{\pm} + \text{jets}$, at the LHC.}
    \label{tab:lhc_lnv_processes}
\end{table}

In the radiative neutrino mass scenario considered here, the flavour structure of the LNV signal is not uniquely determined by neutrino oscillation data. Unlike tree-level scenarios such as the Type-II seesaw model, where the neutrino mass matrix and collider observables are controlled by the same Yukawa structure, the neutrino mass matrix shown in Eq.~\eqref{eqn:solve_for_yukawa} involves two different Yukawa matrices, $\lambda$ and $\lambda^\prime$. In our analysis we assume the relation $\lambda_{ij}=c\lambda'_{ij}$, which reduces the number of free parameters but does not lead to unique predictions for flavour-specific collider observables. 

Furthermore, the flavor composition of the LNV signal depends on the benchmark parameters entering the neutrino mass fit, including the leptoquark masses, mixing angle, and the proportionality constant $c$. Hence, different benchmark points that are equally consistent with neutrino oscillation data can lead to different flavor compositions of the $pp\to \ell^\pm \ell^{\prime\pm}+\text{jets}$ signal. Since the goal of the present work is to establish the inclusive LNV discovery potential at the HL-LHC, we focus on the flavor-inclusive signal rather than flavor-dependent observables.

\section{Analysis}
\label{sec:analysis}

The various leptoquark production modes and decay channels leading to the LNV signature $pp \to \ell^{\pm}\ell'^{\pm} + \text{jets}$ at the LHC are summarized in~Table~\ref{tab:lhc_lnv_processes}. We simulate these LNV signal events using \textsc{MadGraph5\_aMC@NLO}~\cite{Alwall:2014hca}, interfaced with \textsc{Pythia8}~\cite{Sjostrand:2007gs} for parton showering and hadronization. The leptoquark variant of the Zee Model is implemented in \textsc{FeynRules}~\cite{Alloul:2013bka}. Detector effects are simulated with \textsc{Delphes3}~\cite{deFavereau:2013fsa}, employing the default HL-LHC detector card.

We compute cross-sections for the LNV signature $pp \to \ell^{\pm}\ell'^{\pm} + \text{jets}$ arising from  the QCD pair production, EW pair production, and single leptoquark production modes at $\sqrt{s}=13~$TeV. To preserve the lepton number information in the final state, we restrict our analysis to regions of parameter space where the visible decay mode of the lightest leptoquark $\xone \to \ell^{-} j$ dominates or is at least comparable to the neutrino-containing decay channel. From the Yukawa Lagrangian in Eq.~\ref{eqn:Yuk_lag_LQ}, this condition translates to $\lambda^\prime \sin\theta_{\rm LQ} \gtrsim  \lambda \cos\theta_{\rm LQ}$. For the production rates, we compute the leading order (LO) parton level cross-sections using \textsc{MadGraph5\_aMC@NLO} and account for higher-order effects through an overall next-to-leading-order (NLO) $K$-factor. The NLO $K$-factor for the QCD pair production channel is obtained by comparing the LO cross-sections with the NLO prediction~\cite{Kramer:2004df}. For the electroweak pair production and single leptoquark production channels, we restrict ourselves to LO predictions. Given that these two channels are typically suppressed by two orders of magnitude relative to the QCD production mode in the relevant parameter regions, as shown later in this section, neglecting their NLO corrections has a negligible effect on our HL-LHC projections.

\begin{figure}[!t]
    \centering
     \includegraphics[width=0.49\linewidth]{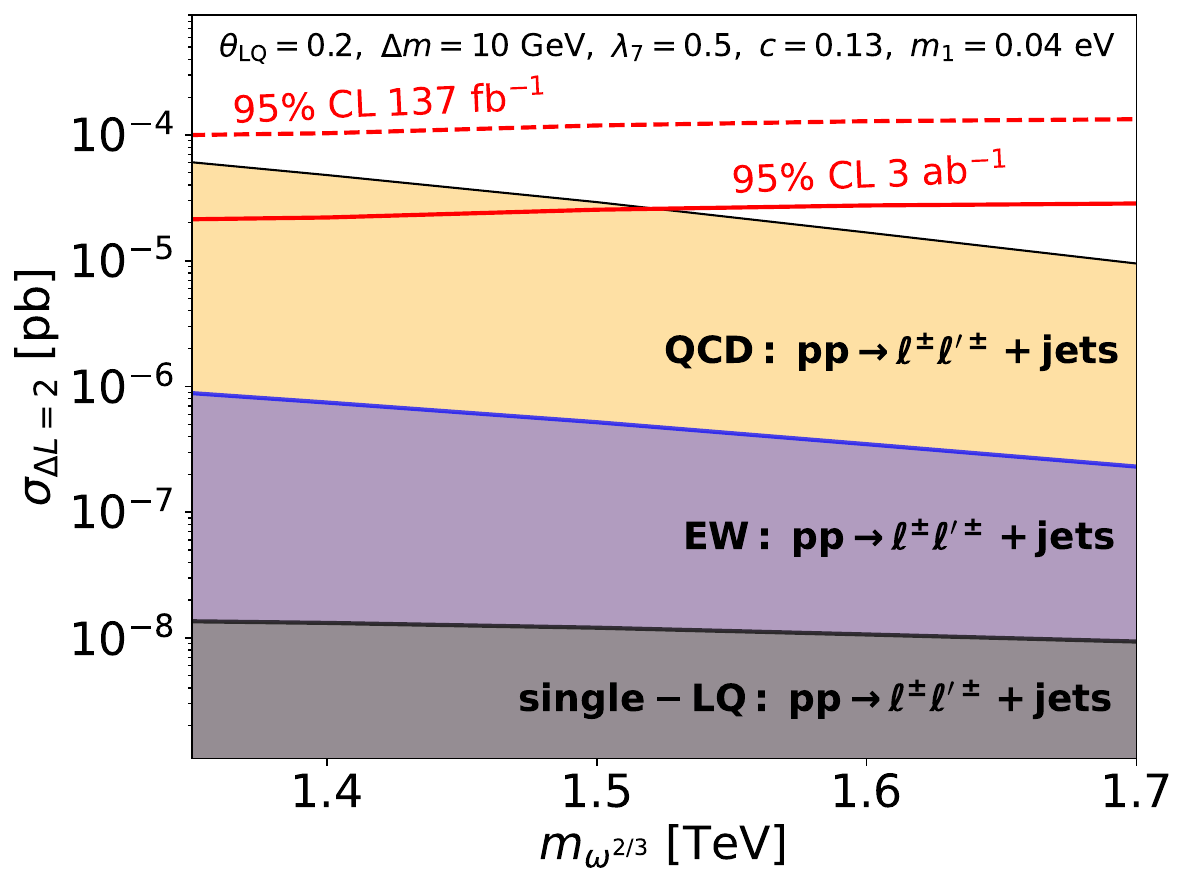}
    \caption{Stacked cross section for the LNV process, $pp \to \ell^{\pm}\ell'^{\pm} + \text{jets}$, mediated by QCD $pp \to \mathrm{LQ~LQ}$~(yellow), EW $pp \to \mathrm{LQ~LQ}$~(blue), and single leptoquark production $pp \to \mathrm{LQ}~\ell$~(grey) at $\sqrt{s} = 13~\text{TeV}$, assuming leptoquark mixing angle $\theta_\text{LQ} = 0.2$, $\Delta m \equiv m_{\omega^{2/3}} - m_{X_1} = 10~\mathrm{GeV}$, $\lambda_7 = 0.5$, $c \equiv \lambda_{ij}/\lambda'_{ij} = 0.13$, and $m_1 = 0.04~\mathrm{eV}$. The 95\% confidence level limits are shown for an integrated luminosity of 137~fb$^{-1}$ (red dashed) and for the HL-LHC projection with 3~ab$^{-1}$ (red solid).}
    \label{fig:stacked_cross-sections}
\end{figure}


In Fig.~\ref{fig:stacked_cross-sections}, we present the stacked cross-sections for $pp \to \ell^{\pm}\ell'^{\pm} + \text{jets}$ mediated by QCD pair production, EW pair production, and single leptoquark production as a function of the leptoquark mass $m_{\om}$. The QCD pair production mode dominates over the entire mass range, yielding cross sections several orders of magnitude larger than those from EW pair or single leptoquark production for both benchmark scenarios. This enhancement arises from the purely QCD-driven nature of that production mode, larger gluon parton distribution functions in the proton, and the sizable color factor associated with gluon-gluon initiated channels. In contrast,  EW leptoquark pair production is suppressed due to their dependence on electroweak couplings and quark-initiated parton luminosities. Single leptoquark production is further reduced, as it requires a leptoquark Yukawa coupling insertion already at the production level.\footnote{This explains why the QCD-driven channel in Fig.~\ref{fig:lnv_feyn} (left panel) yields the largest rate among the three LNV production mechanisms, even though the corresponding neutrino-mass diagram arises only at higher loop order, as shown in Fig.~\ref{fig:lnumass_feyn}.} For instance, the $qg \to \omstar \ell^+$ channel is suppressed by a factor of $\sim |\lambda|^2$. In the allowed parameter space, the Yukawa couplings to the valence quarks are typically small $|\lambda_{ed}|,|\lambda_{\mu d}| \lesssim 10^{-3}$, resulting in subdominant contributions from the single leptoquark channel, as observed in Fig.~\ref{fig:stacked_cross-sections}. Despite this suppression, we account for all three production mechanisms in our signal simulation for completeness.

It is worth emphasizing that a sizable LNV signal in this model does not imply an enhanced neutrino mass contribution. The neutrino mass matrix is generated radiatively and is constrained to reproduce the observed neutrino masses and mixing parameters through Eq.~\eqref{eqn:solve_for_yukawa}. By contrast, the dominant contribution to the LNV signal at the LHC arises from QCD-mediated leptoquark pair production, whose cross section depends primarily on the leptoquark masses and is largely independent of the Yukawa couplings entering the neutrino mass matrix. The Yukawa couplings mainly affect the subsequent decay branching fractions of the leptoquarks. Thus, an observable LNV signal at the HL-LHC is fully compatible with the small neutrino masses required by current neutrino oscillation data.

The LNV signal $pp \to \ell^{\pm}\ell'^{\pm} + \text{jets}$ is subject to substantial background arising from non-prompt leptons, which are challenging to simulate reliably. Therefore, we adopt the background estimates reported in the CMS same-sign dilepton plus jets search with $137~\mathrm{fb}^{-1}$ of data~\cite{Sirunyan:2020ztc}, and rescale them to $3~\mathrm{ab}^{-1}$ for HL-LHC projections. Following the CMS event selection criteria in \cite{Sirunyan:2020ztc}, we require signal events to contain exactly two same-sign isolated leptons with transverse momentum $p_{T\ell} > 25~\mathrm{GeV}$ and pseudorapidity $|\eta_e| < 2.5$ for electrons and $|\eta_\mu| < 2.4$ for muons. Events must also include at least two jets, reconstructed using the anti-$k_T$ jet algorithm with radius $R = 0.4$, satisfying $p_{Tj} > 40~\mathrm{GeV}$ and $|\eta_j| < 2.4$. To suppress background contributions, events with same-flavor dileptons having invariant mass $m_{\ell\ell} < 12~$GeV, or different-flavor leptons with $m_{\ell\ell'} < 8$~GeV, are vetoed. Events are also required to have only small missing energy $\slashed{E}_{T} < 50~$GeV and a large scalar sum of the transverse momenta $H_T > 1300~$GeV. 

Using these event selection criteria and background estimates from the CMS analysis in Ref.~\cite{Sirunyan:2020ztc}, we evaluate the projected sensitivity to LNV at the HL-LHC. To illustrate the potential reach, we consider a representative benchmark point in the currently allowed parameter space:  $\theta_\mathrm{LQ} = 0.2$, $m_{\omega^{2/3}}=1.4$~TeV, $\Delta m = 10~\mathrm{GeV}$, $\lambda_7 = 0.5$, $c \equiv \lambda_{ij}/\lambda'_{ij} = 0.13$, and $m_1 = 0.04~\mathrm{eV}$. 
The signal significance is quantified using $S/\sqrt{B}$, where $S$ is the expected number of signal events and $B$ is the luminosity-scaled projected background rate derived from the CMS analysis~\cite{Sirunyan:2020ztc}. 

\begin{figure}[!t]
    \centering
    \includegraphics[width=0.49\linewidth]{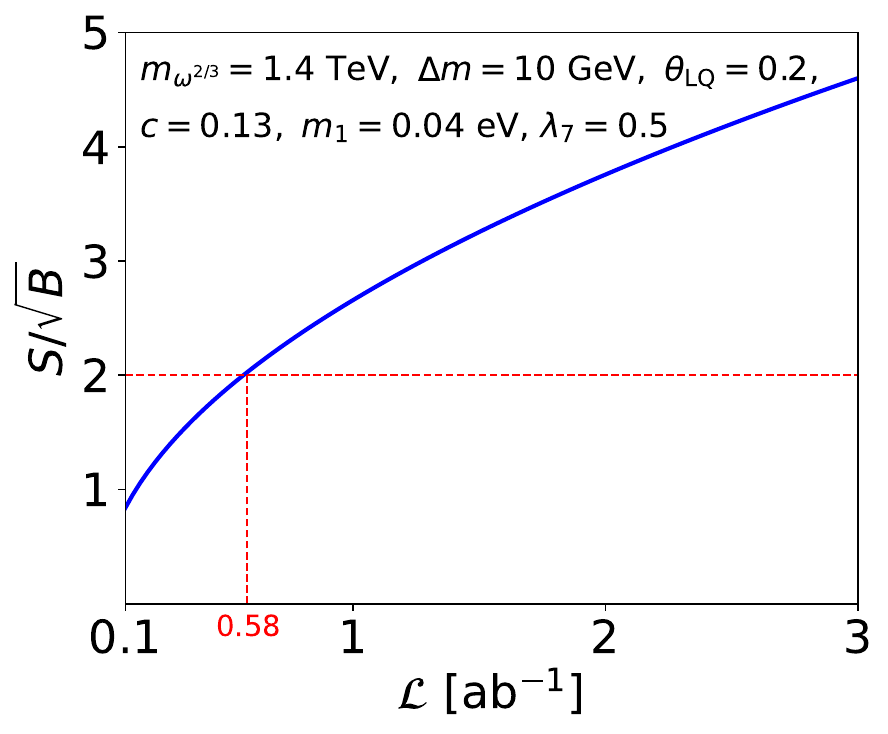}\includegraphics[width=0.48\linewidth]{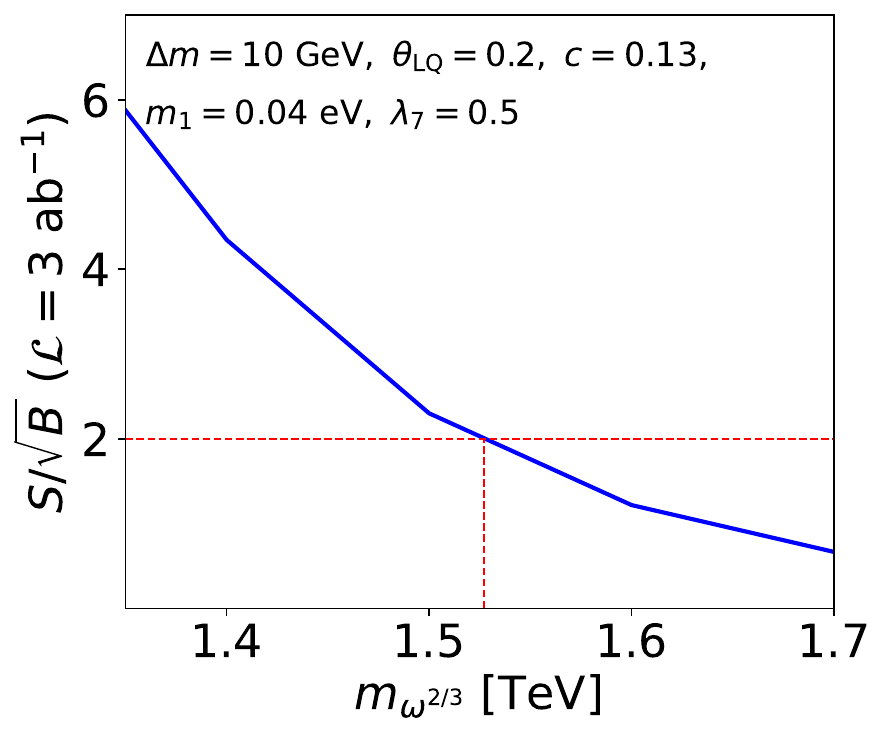}
    \caption{Signal significance $S/\sqrt{B}$ for the LNV process $pp \to \ell^{\pm}\ell'^{\pm} + \text{jets}$ at the LHC with $\sqrt{s} = 13~\mathrm{TeV}$. Left: Significance as a function of integrated luminosity. Right: Significance at the HL-LHC with $\mathcal{L}=3~\text{ab}^{-1}$  as a function of the leptoquark mass $m_{\om}$, assuming $\theta_\text{LQ} = 0.2$, $\Delta m = 10~\mathrm{GeV}$, $\lambda_7 = 0.5$, $c \equiv \lambda_{ij}/\lambda'_{ij} = 0.13$, and $m_1 = 0.04~\mathrm{eV}$.}
    \label{fig:significance_vs_luminosity}
\end{figure}

In Fig.~\ref{fig:significance_vs_luminosity} (left panel), we show the signal significance as a function of the integrated luminosity $\mathcal{L}$. For the representative parameter choices, a $2\sigma$ sensitivity to LNV can be achieved around $\mathcal{L} \sim 790~\mathrm{fb}^{-1}$. In Fig.~\ref{fig:significance_vs_luminosity}~(right panel), the signal significance is shown as a function of leptoquark mass $m_{\om}$, assuming a mass splitting $\Delta m = m_{\om} - m_{\xone} = 10~$GeV. We observe that the HL-LHC with $\mathcal{L}=3~\text{ab}^{-1}$  will be able to probe the LNV signal for leptoquark masses up to approximately $m_{\om} \sim 1.5$~TeV.

\section{Conclusion}
\label{sec:conclusion}

Determining whether lepton number is a conserved symmetry in nature remains a fundamental question, closely linked to the Majorana versus Dirac character of neutrinos. Observation of lepton number violation by two units at colliders would provide a direct and complementary probe to neutrinoless double beta decay, opening a window into the mechanism behind neutrino mass generation. Motivated by this, we have investigated the potential for detecting the LNV process  $pp \to \ell^{\pm}\ell'^{\pm} + \text{jets}$ at the HL-LHC within the leptoquark variant of the Zee Model.

In this scenario, neutrino masses are generated radiatively at one-loop, while the same interactions give rise to distinct LNV signatures at colliders. After applying current experimental constraints, including neutrino mass bounds, charged lepton flavor violation limits, electroweak precision tests, and existing LHC searches for leptoquarks, we identified the allowed parameter space and the dominant production mechanisms for the $|\Delta L|=2$ final state. We divide these channels into three classes: QCD leptoquark pair production, electroweak pair production, and single leptoquark production. Among these, QCD pair production overwhelmingly dominates the signal across the relevant parameter space.

Based on background estimates from the CMS same-sign dilepton plus jets analysis~\cite{Sirunyan:2020ztc}, we performed a collider sensitivity study for the HL-LHC. Our results indicate that leptoquark masses up to $m_{\rm LQ} \sim 1.5~\mathrm{TeV}$ could be probed through this LNV channel. These results demonstrate that the same-sign dilepton plus jets signature is a promising probe of lepton number violation at the HL-LHC, providing a direct test of leptoquark-induced radiative neutrino mass generation mechanism.

\section*{Acknowledgments}
The work of KSB and DG is supported by the U.S. Department of Energy under grant number DE-SC0016013. The work of RKB is supported by the World Premier International Research Center Initiative (WPI), MEXT, Japan, and by JSPS KAKENHI Grant Number JP24K22876. This work used computing resources at Kavli IPMU.

\providecommand{\href}[2]{#2}\begingroup\raggedright\endgroup

\end{document}